\pdfoutput=1 
\documentclass{JINST}

\usepackage{subfigure}

\usepackage{lineno}

\title{Performance of Photosensors in the PandaX-I Experiment}

\author{Shaoli Li$^a$\thanks{Corresponding author, lisl@sjtu.edu.cn.},
Xun Chen$^a$,
Karl L. Giboni$^a$,
Guodong Guo$^a$,
Xiangdong Ji$^{a,b,c}$,
Qing Lin$^a$,
Jianglai Liu$^a$\thanks{Corresponding author, jianglai.liu@sjtu.edu.cn.},
Yajun Mao$^b$, 
Kaixuan Ni$^a$\thanks{Now at Department of Physics, University of California, San Diego.},
Xiangxiang Ren$^{d,a}$,
Andi Tan$^c$,
Mengjiao Xiao$^a$,
Xiang Xiao$^a$,
Xiaopeng Zhou$^b$\\
\llap{$^a$}INPAC and Department of Physics and Astronomy, Shanghai Jiao Tong University, \\
Shanghai Key Laboratory for Particle Physics and Cosmology, Shanghai, 200240, China\\
\llap{$^b$}School of Physics, Peking University, Beijing, 100080, China\\
\llap{$^c$}Department of Physics, University of Maryland, College Park, MD, 20742, USA\\
\llap{$^d$}School of Physics and Key Laboratory of Particle Physics and Particle Irradiation (MOE), Shandong University, Jinan 250100, China\\
}

\abstract{
We report the long term performance of the photosensors, 
143 one-inch R8520-406 and 37 three-inch 
R11410-MOD photomultipliers from Hamamatsu, in the first 
phase of the PandaX dual-phase xenon dark matter experiment. This is the first time 
that a significant number of R11410 
photomultiplier tubes were operated in liquid xenon for an extended period, providing 
important guidance to the future large xenon-based dark matter experiments.  
}

 
\begin{document}


\section{Introduction}
\label{}
The PandaX project is a series of deep 
underground experiments aiming to study the properties of dark matter and 
neutrinos using xenon detectors in the China Jin-Ping Underground Laboratory~\cite{CJPL}. 
The first phase of the experiment, 
PandaX-I~\cite{pandaxI}, is a dual-phase xenon dark matter experiment using 
a time projection chamber (TPC), a  technique similar to that 
used in earlier XENON10~\cite{xenon10}, ZEPLIN-III~\cite{zeplin}, 
XENON100~\cite{xenon100}, 
and LUX~\cite{lux} experiments. The TPC 
contains an active target of $\sim$120 kg of liquid xenon. 
A stainless steel inner vessel is used as the liquid xenon cryostat, enclosed 
by a copper vacuum chamber acting as 
a vacuum jacket for the cryostat, a shield for the external gamma rays, 
as well as a radon barrier to the detector. 
The liquid xenon target is enclosed by a cylindrical TPC 
consisting of a field cage to drift the ionization electrons in the liquid 
and to create proportional scintillation in the 
gas for those electrons via electroluminescense. A
top and bottom UV (178 nm) sensitive photomultiplier tube (PMT) arrays made up of 
143 Hamamatsu R8520-406 one-inch 10-stage PMTs and 
37 Hamamatsu R11410-MOD high quantum efficiency 
three-inch 12-stage PMTs, respectively, 
are looking into the target 
collecting photons from the prompt scintillation signals (S1) with a width of 
$\sim$100 ns and the delayed ionization signals (S2) with a width of 
$\sim$2~$\mu$s. The detected light pattern on the PMTs as well as the 
time separation between the S1 and S2 allow reconstructions of the location of the 
energy deposition. The wall of the field cage is made out of polytetrafluoroethylene (PTFE) 
reflective 
panels to enhance the light collection. 

PandaX-I started its liquid xenon operation in February 2014, and 
concluded data taking in October 2014, with an overall 80.1 live-day dark matter search
data collected and published~\cite{pandaxI1st, pandaxI2nd}. Here we report
the PMT performance during the entire data taking period. 
The remainder of this paper is organized as follows. In Sec.~\ref{sec:base}, 
we briefly describe the PMT assemblies, readout
connections and signal processing. Then 
in Sec.~\ref{sec:performance} we discuss details of the PMT performance 
within the nine months of cryogenic operation, with some 
emphasis on the R11410-MOD tubes (bottom PMTs) since
they are used for the first time with large quantities in a liquid-xenon-based experiment,
followed by a short summary at the end. 

\section{PMT assemblies, connections, and signal processing}
\label{sec:base}
\begin{figure}[!htbp]
\centering
\includegraphics[width=0.8\linewidth]{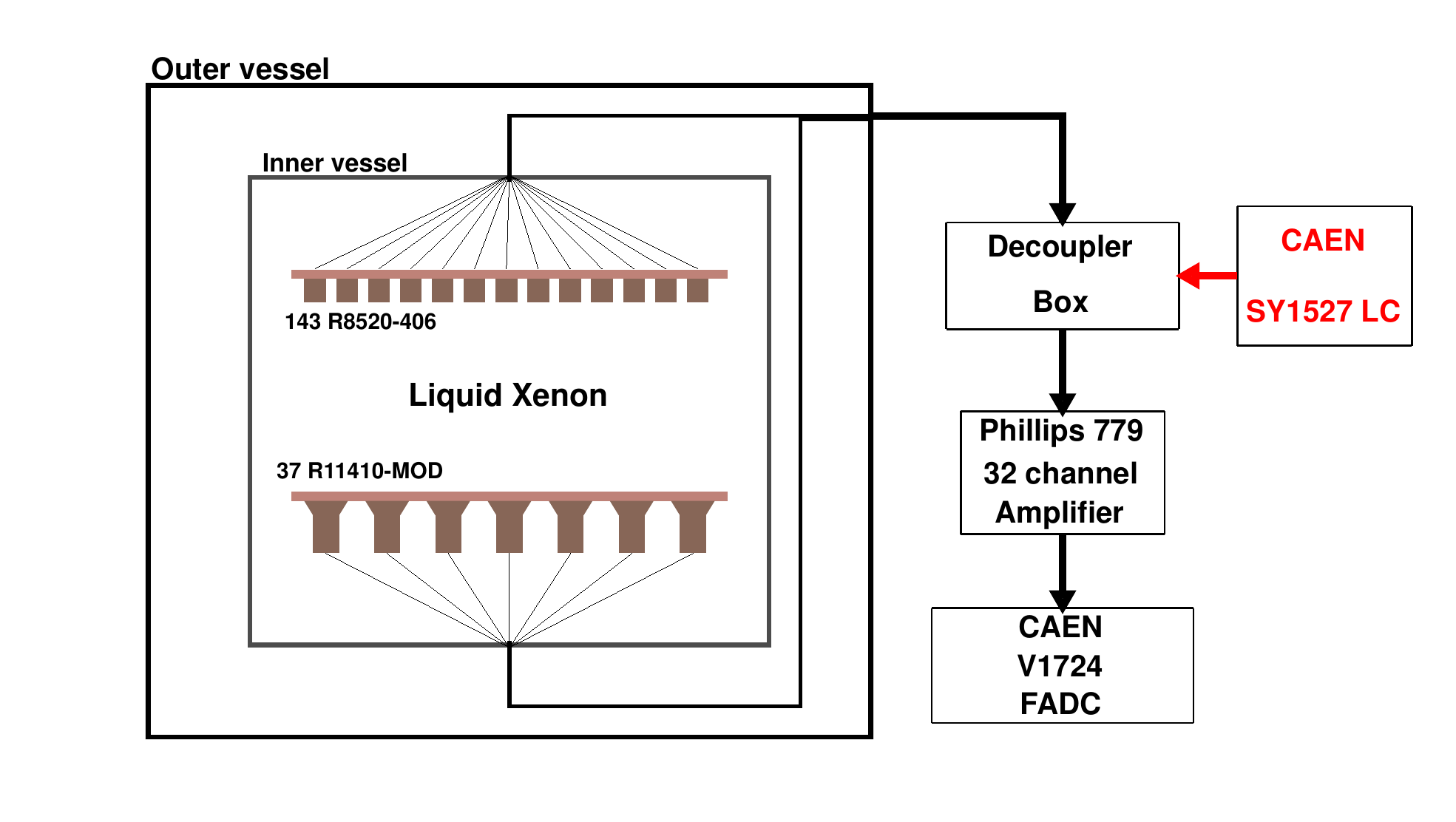}
\caption{A diagram illustrating the connections between the PMTs
and their readout electronics.}
\label{fig:pandaxI_overview}
\end{figure}

A diagram illustrating the connection between the PMT assemblies and the 
readout electronics is shown in Fig.~\ref{fig:pandaxI_overview}.
For each PMT, a positive high voltage (HV) divider circuit, 
customly built with low radioactivity Cirlex PCB, 
was implemented with positive HV and signal combined on 
a single output miniature coaxial cable
to minimize the number of cables in the 
cryostat. The design of the PMT bases was discussed in Ref.~\cite{pandaxI}. 
The voltage divider followed Hamamatsu's recommended divider ratio. The overall 
base resistance was 12.5 and 18.5 M$\Omega$, respectively, for R8520-406 and R11410-MOD. 
The back-termination resistor at the anode output was chosen to be 100 k$\Omega$ to 
increase the low frequency band width in order to avoid distortion of the S2 signals. 
To minimize radioactivity due to ceramic capacitors, 
we only kept one 10 nF capacitor between the last dynode and the anode, since 
the linearity was not a concern at the low energy dark matter signal region. 

The cables for the top and bottom arrays were fed out through the feedthroughs located 
at the top and bottom of the inner vessel, respectively, then through a second 
feedthrough manifold located on the outer vessel to the electronics area.
The HV and signal were decoupled using a three-stage RC decoupling
circuit~\cite{pandaxI}. The value of the decoupling capacitor was chosen to be 100 nF
in order to minimize the signal distortion for S2s. 
The PMT signals were amplified by the 
Phillips 779 linear amplifiers, then input into the CAEN V1724 digitizers to be read out 
by the data acquisition. 


\section{PMT performance}
\label{sec:performance}
During the experiment, the PMTs were calibrated weekly using LEDs.
Three LEDs (390 to 395 nm) outside the detectors were 
driven by a standard driver circuitry~\cite{paper:lanlLED} to produce fast ($\sim$10~ns) 
and feeble
light pulses, which were transmitted into the detector through three optical fibers 
and were further distributed by three Teflon diffuser rods mounted outside the 
PTFE panels. The driver also provided 
synchronized triggers for the data acquisition. The voltage of the driver was adjusted 
to ensure $<$20\% occupancies for every PMT in order to cleanly separate the 
single photoelectrons (SPEs). In addition, the gains, random PMT rates, 
as well as afterpulsing rates were monitored {\it in situ} 
during normal data taking. In this 
section, a few key properties of the PMTs shall be discussed in turn. 

\subsection{Gain}
\begin{figure}[!htbp]
\centering
  \includegraphics[width=0.7\linewidth]{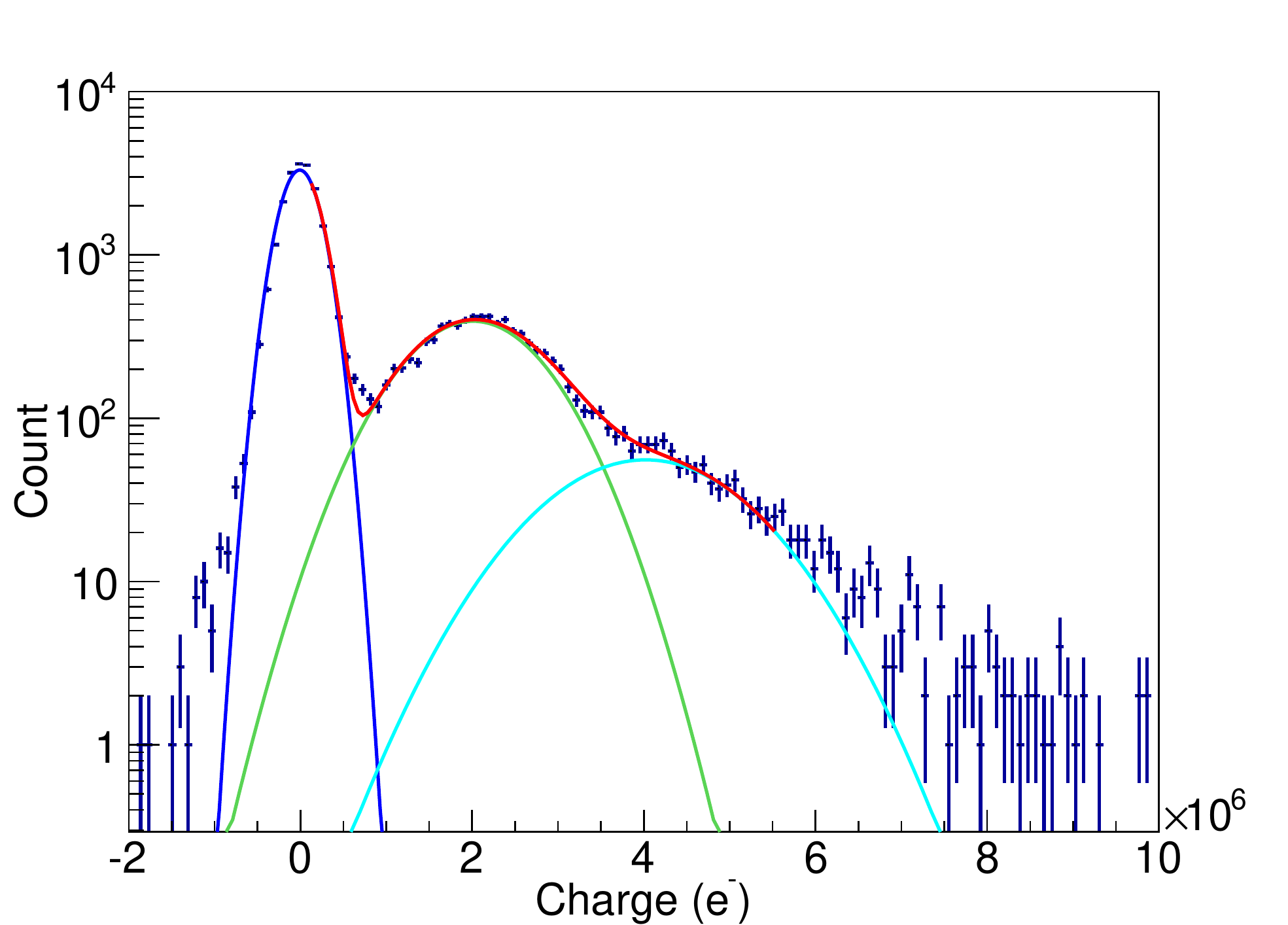}
  \caption{Low occupancy charge spectrum for a typical R11410-MOD PMT 
    with a gain of $\sim2\times10^6$ during the 
    LED calibration run in liquid xenon, 
    and a combined fit of the pedstal (blue), SPE (green), and double PE (cyan)
    with the fit range indicated by the fit curve (red).
  }
  \label{fig:spe_spectrum}
\end{figure}

\begin{figure}[!htbp]
\centering
\subfigure[\it R11410-MOD]
{
  \includegraphics[width=.45\linewidth]{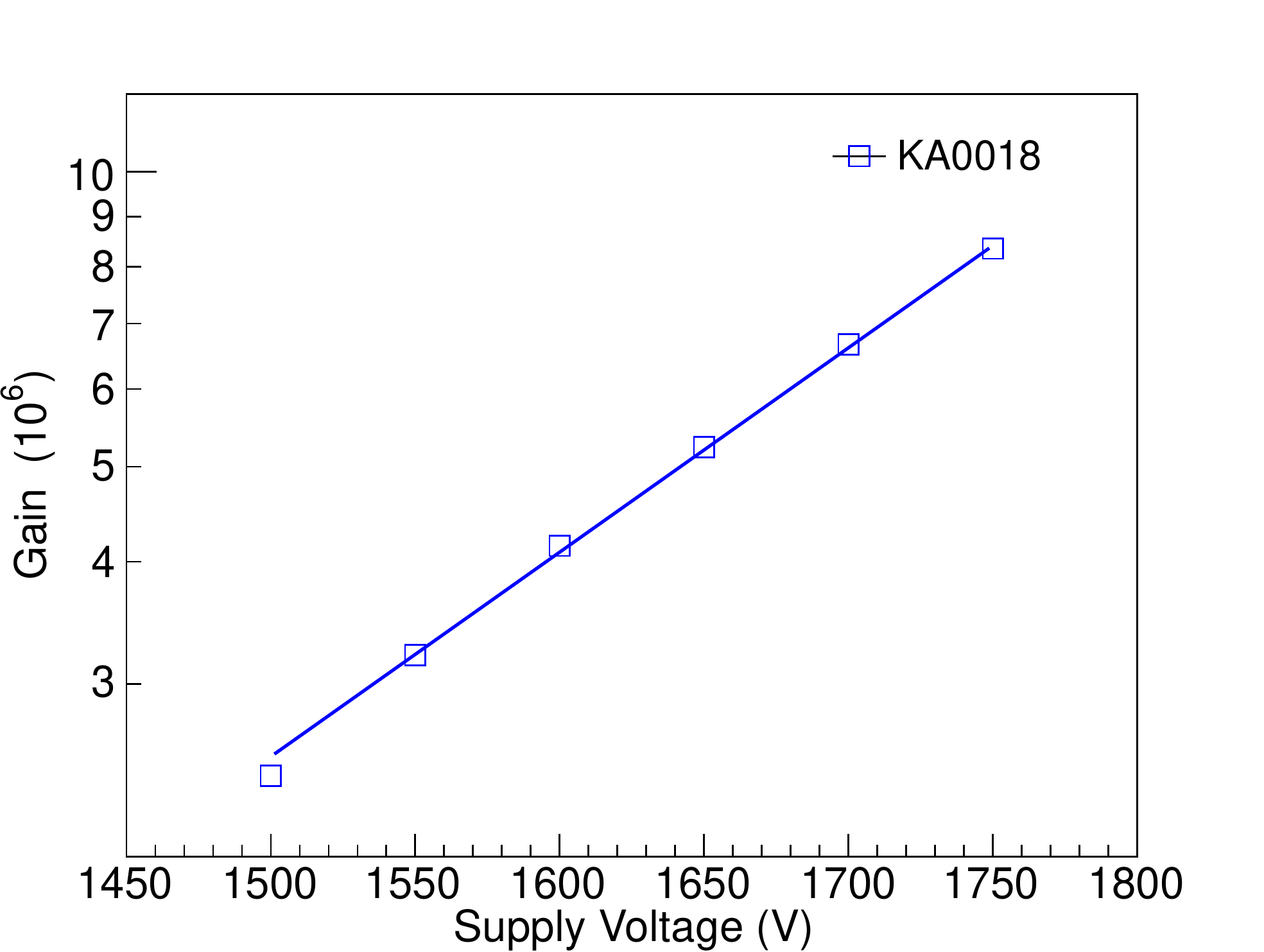}
}
\subfigure[\it R8520-406]
{
  \includegraphics[width=.45\linewidth]{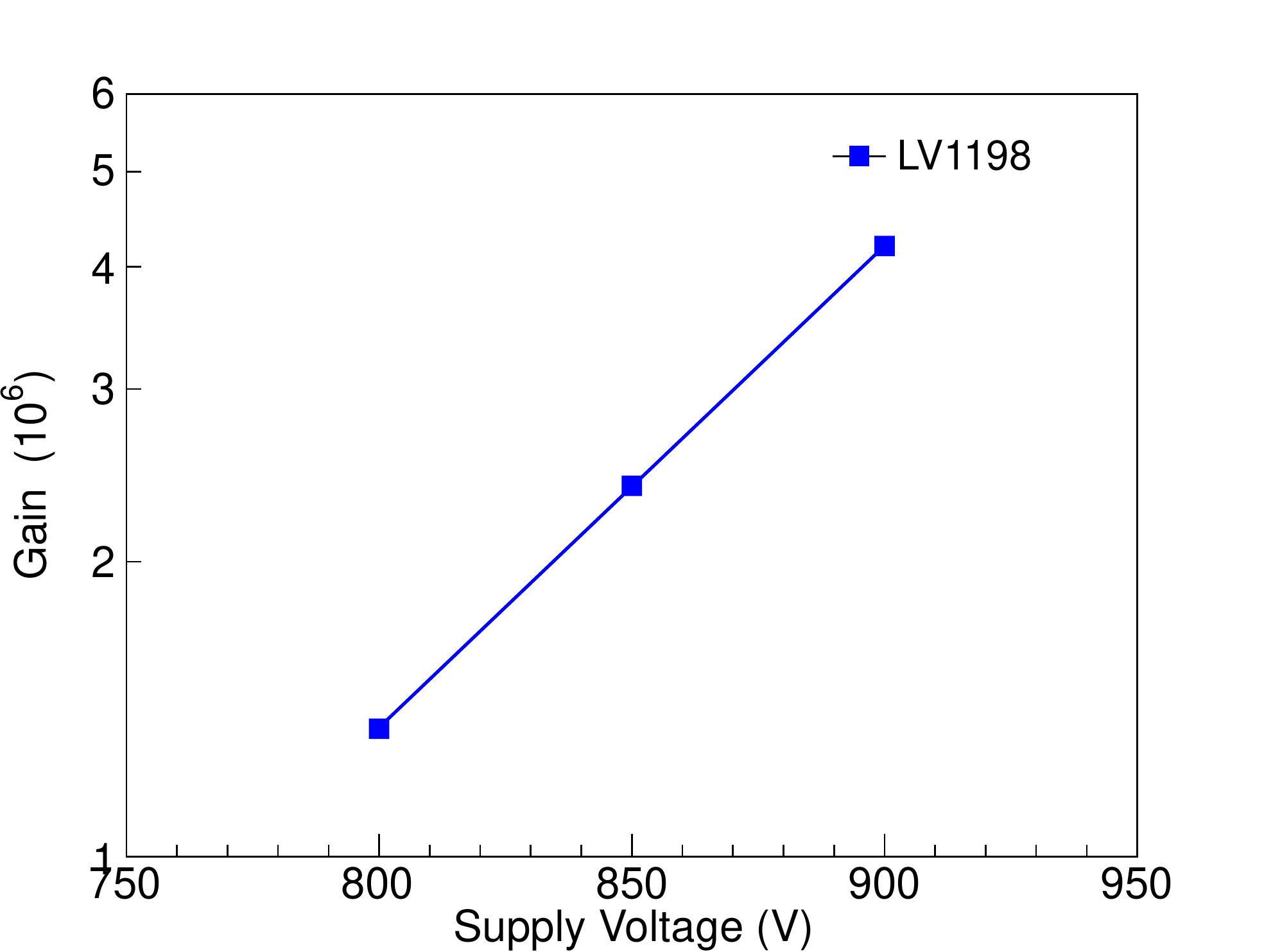}
}
\caption{Gain vs. supply voltage for a typical R11410-MOD (a) and R8540-406 (b) 
PMT, showing an expected exponential relation with the supply voltage.}
\label{fig:gain_hv}
\end{figure}

The area of SPE waveform (in units of $e$) 
was used to calibrate the gain of each PMT. 
A typical charge spectrum for a R11410-MOD PMT is shown in 
Figure~\ref{fig:spe_spectrum}.
The measured PMT charge distribution $f(q)$ is fitted with
\begin{equation}
\label{eq:spe_func}
f(q) =  c_{1}\times G(q, \mu_{1},\sigma_{1}) + c_{2}\times G(q, \mu_{2}+\mu_1, \sqrt{\sigma_{2}^2+\sigma_{1}^2}) + c_{3}\times G(q, 2\mu_{2}+\mu_1,\sqrt{2\sigma_{2}^2+\sigma_1^2})\,,
\end{equation}
where $G(x, \mu, \sigma) = e^{-(x-\mu)^2/2\sigma^2}$ is a standard Gaussian function. 
The first term describes the pedestal which comes from the electronic noise
with a centroid and width of $\mu_1\sim0$ and $\sigma_1\sim0.3\times10^6$, respectively.
Similarly the second and third terms 
represent the single and double photoelectrons peaks,  respectively, with their 
centroids and widths properly related. 
The SPE gain is typically $\sim$2$\times$10$^{6}$. 
Typical gain values at different supply voltages for the two types of 
PMTs are shown in Fig.~\ref{fig:gain_hv}, showing expected exponential relations.
To operate under a uniform normal gain of 2$\times$10$^6$, 
the supply voltage was set ranging from 1264 V to 1550 V (R11410-MOD) and 
675 V to 881 V (R8520-406), respectively, for all the PMTs. 

\begin{figure}[!htbp]
\centering
  \includegraphics[width=0.8\linewidth]{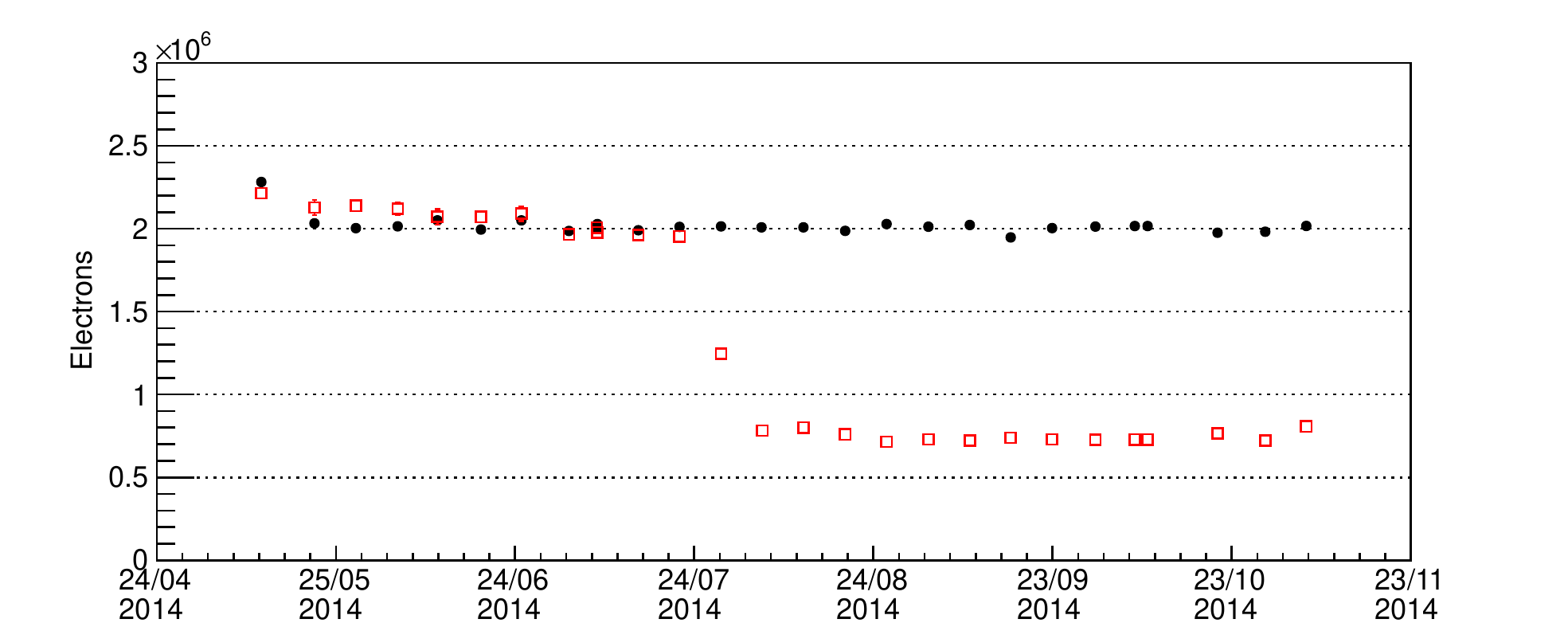}
  \caption{The gain history of two R11410-MOD tubes. Black dots are from a typical
    stable PMT. Red squares represent the gain of a PMT with the supplied 
    voltage lowered twice during the run
    (from the 11$^{th}$ to 12$^{th}$, and 12$^{th}$ to 13$^{th}$ points)
    to avoid excessive random PMT rates.}
  \label{fig:gain_history}
\end{figure}

In Fig.~\ref{fig:gain_history}, the stability of the gains of two R11410-MOD PMTs 
during six months stable data taking is shown. Most of the PMT gains are 
stable to $<$10\% except 
a few that we had to lower the supply voltage in order to avoid excessive random PMT 
rates  (see Sec.~\ref{sec:dark_rate}). 

\begin{figure}[!htbp]
\centering
\subfigure[\it R11410-MOD]
{
  \includegraphics[width=.45\linewidth]{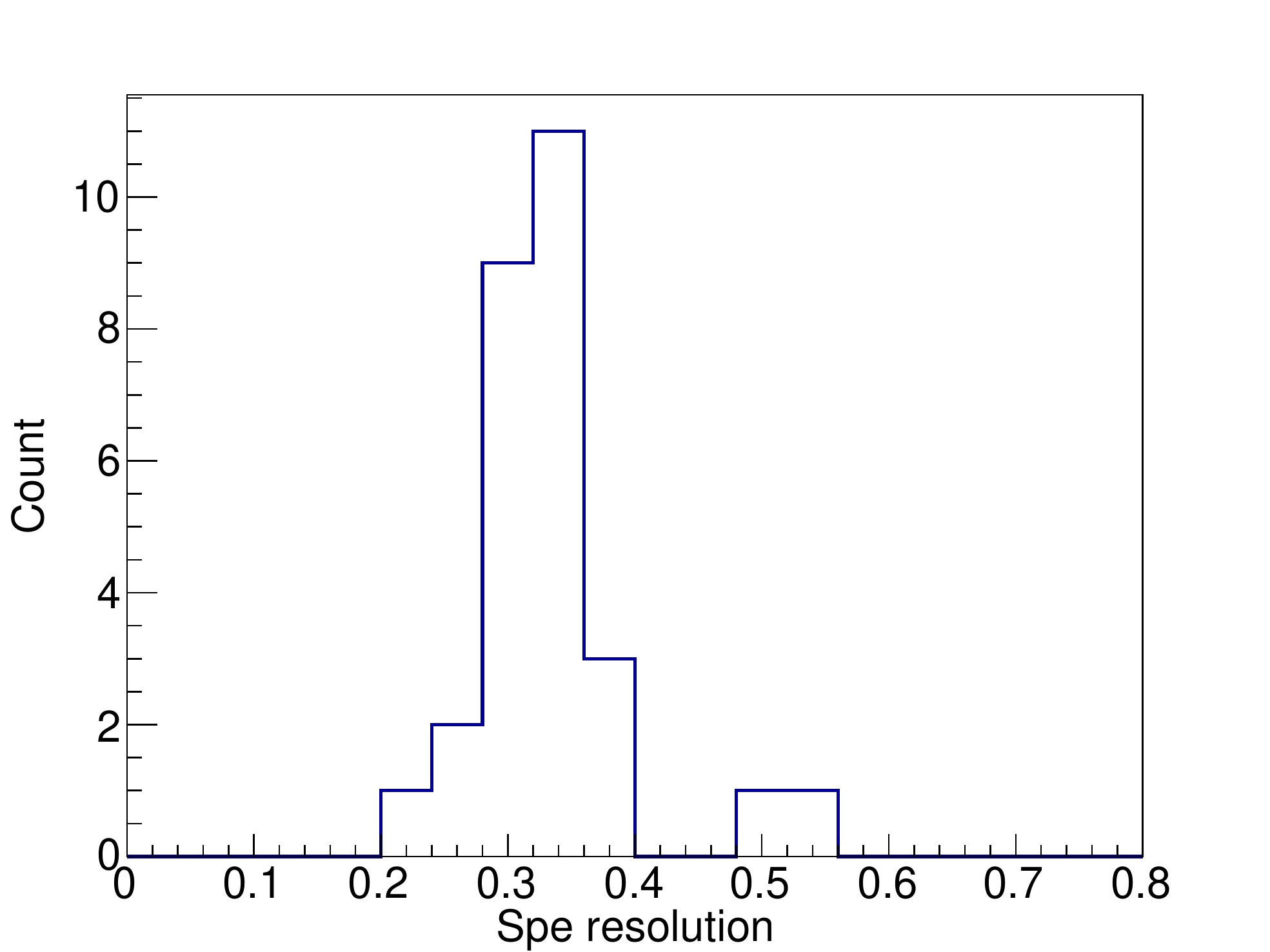}
}
\subfigure[\it R8520-406]
{
  \includegraphics[width=.45\linewidth]{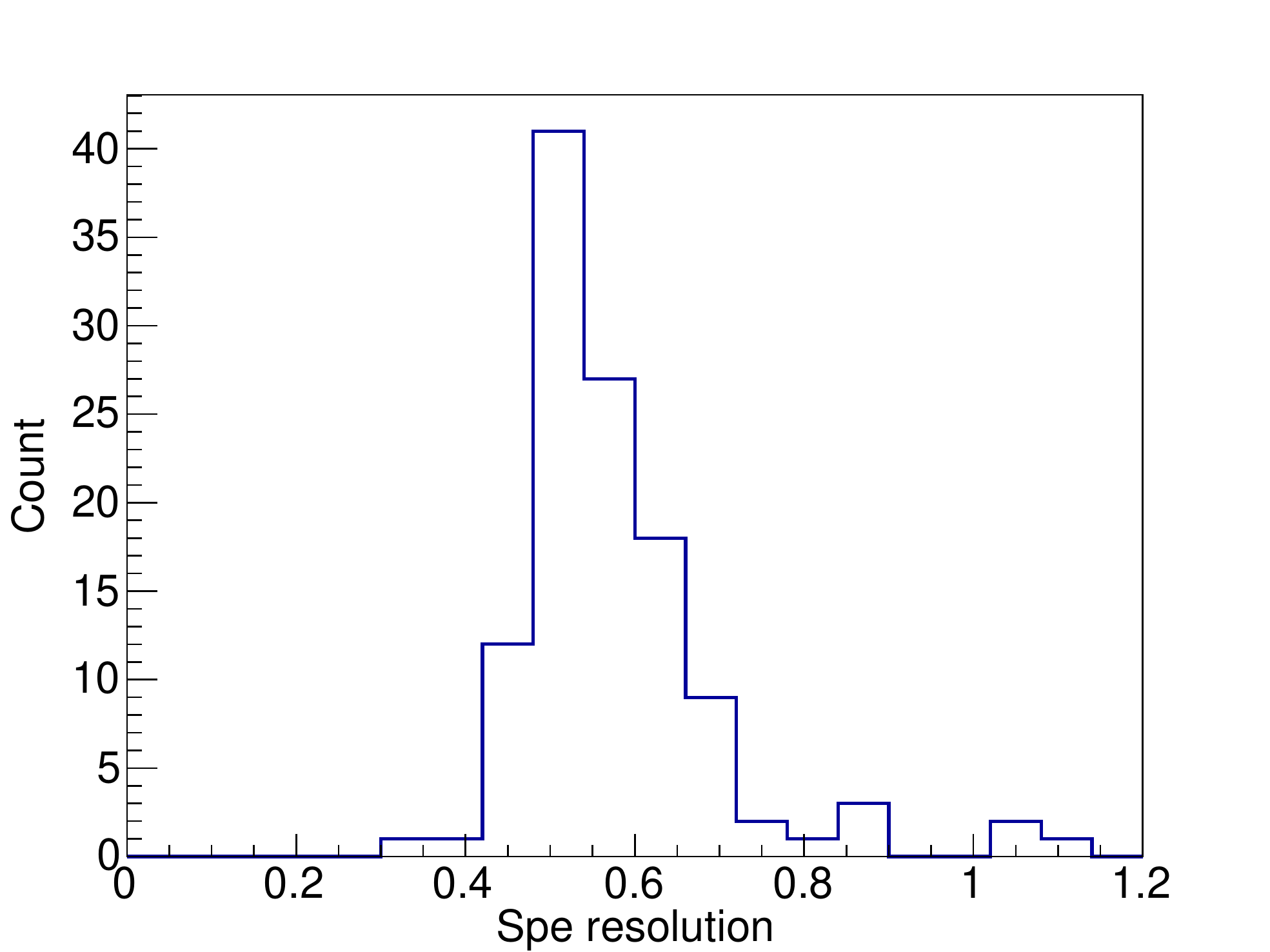}
}
\caption{SPE resolutions of all R11410-MOD (a) and R8520-406 (b) PMTs 
operated under normal gain obtained in a typical calibration run.}
\label{fig:gain_spe_ptv}
\end{figure}

The resolution of a PMT to single photoelectrons was obtained from the LED 
calibration data. It is defined as $\sigma_2/\mu_2$ based on Eqn.~\ref{eq:spe_func}.
For all the PMTs working stably under the normal gain,  
the SPE resolutions are summarized in Fig.~\ref{fig:gain_spe_ptv}. The average is
35\% for R11410-MOD and 58\% for R8520-406. 

\subsection{Random PMT rate}
\label{sec:dark_rate}
\begin{figure}[!htbp]
\centering
\subfigure[\it R11410-MOD]
{
  \label{fig:dark_rate_R11410}
  \includegraphics[width=0.45\linewidth]{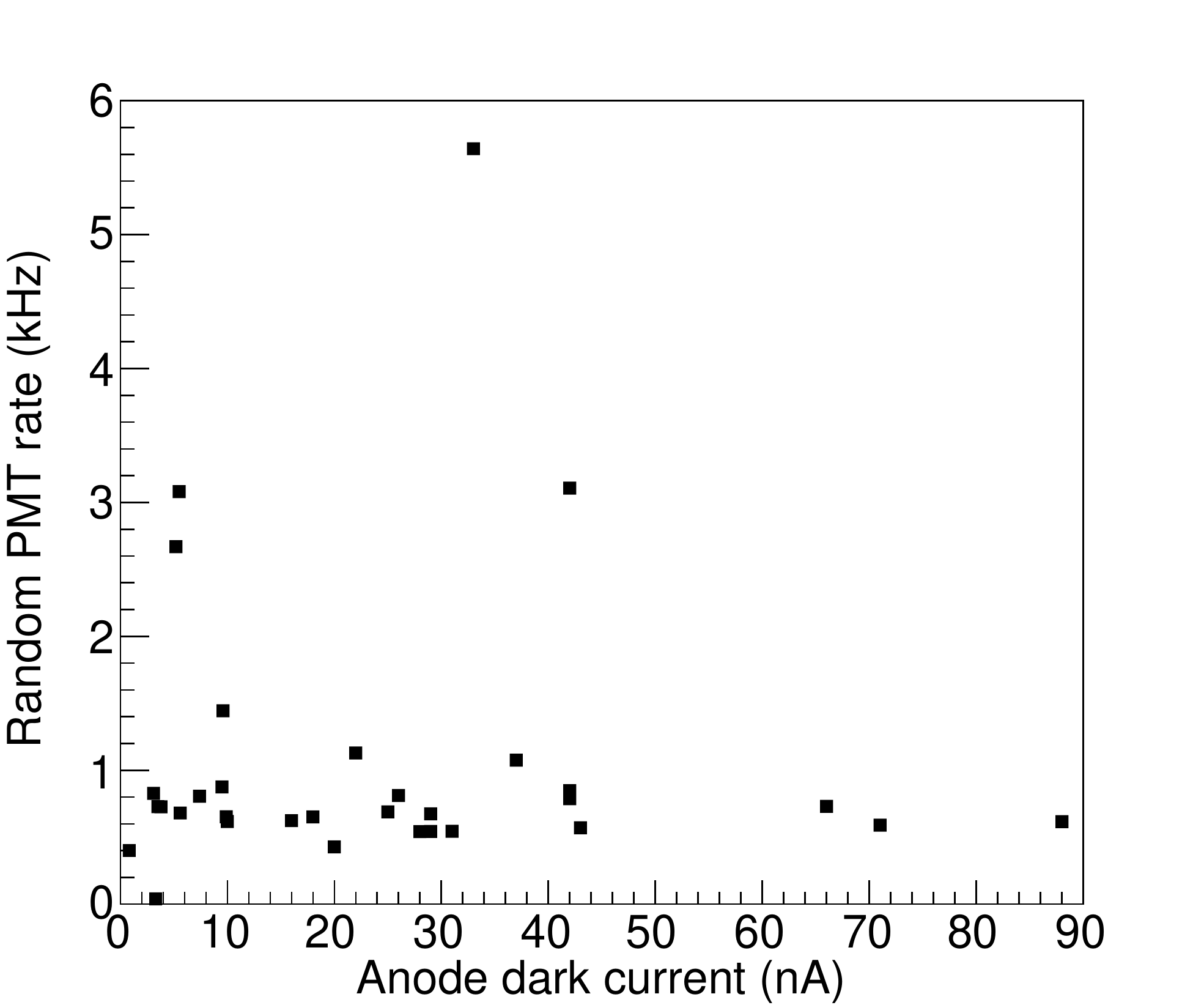}
}
\subfigure[\it R8520-406]
{
  \label{fig:dark_rate_R8520}
  \includegraphics[width=0.45\linewidth]{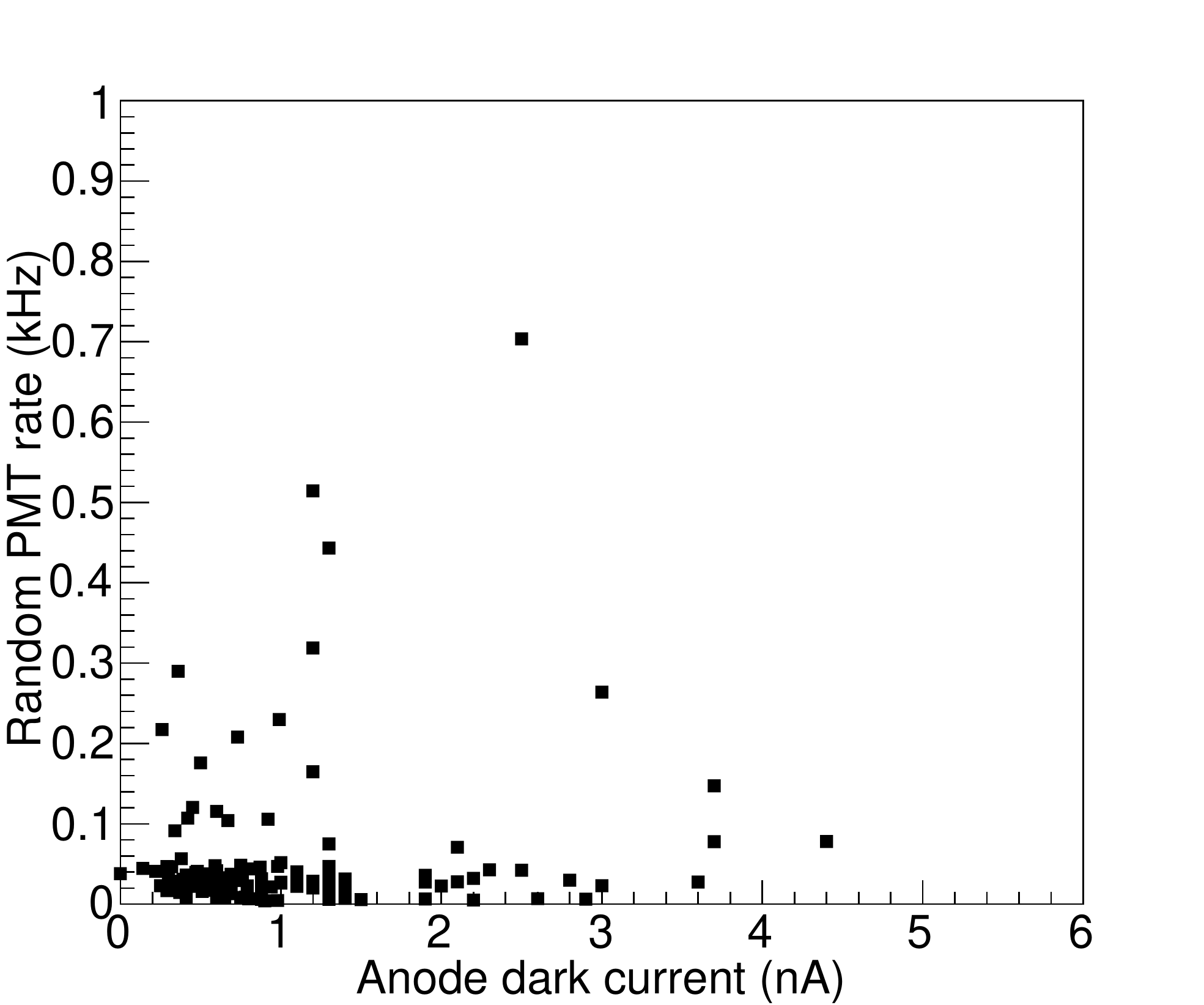}
}
\caption{Random PMT rate vs. anode dark current from Hamamatsu data sheets.
  }
\end{figure}

We define the random PMT rate as the rate of random SPE-like hits of each 
PMT independent of
the physical event triggers.
In the regular running of PandaX-I, each full data acquisition window is
200 $\mu$s with 100 $\mu$s before the trigger.
The maximum drift time in the 15 cm liquid xenon is 88 $\mu$s (1.7 mm/$\mu$s drift
velocity)~\cite{pandaxI}. We calculate the random PMT rate based on the number of SPE-like
pulses in the first 5 $\mu$s at the beginning of the waveform where no genuine S1 signal
is expected.
In Figs.~\ref{fig:dark_rate_R11410} and~\ref{fig:dark_rate_R8520}, the
average random PMT rates during the liquid xenon run vs. dark currents 
from the Hamamatsu data sheet are shown for R11410-MOD and R8520-406, respectively. 
On average, the random PMT rates were 1.07 kHz (R11410-MOD) and 0.06 kHz (R8520-406) 
during the operation in liquid xenon, whereas those measured 
in gaseous xenon under room temperature were 1.3 kHz and 0.3 kHz, respectively. 
In addition, no clear correlation between the random PMT rate 
and dark current was observed, indicating that the former 
might be dominated by additional light generation in the 
liquid xenon detector instead of the intrinsic thermionic electron emission and 
leakage current from the PMTs measured by the manufacturer.

\begin{figure}[!htbp]
\centering
  \includegraphics[width=0.8\linewidth]{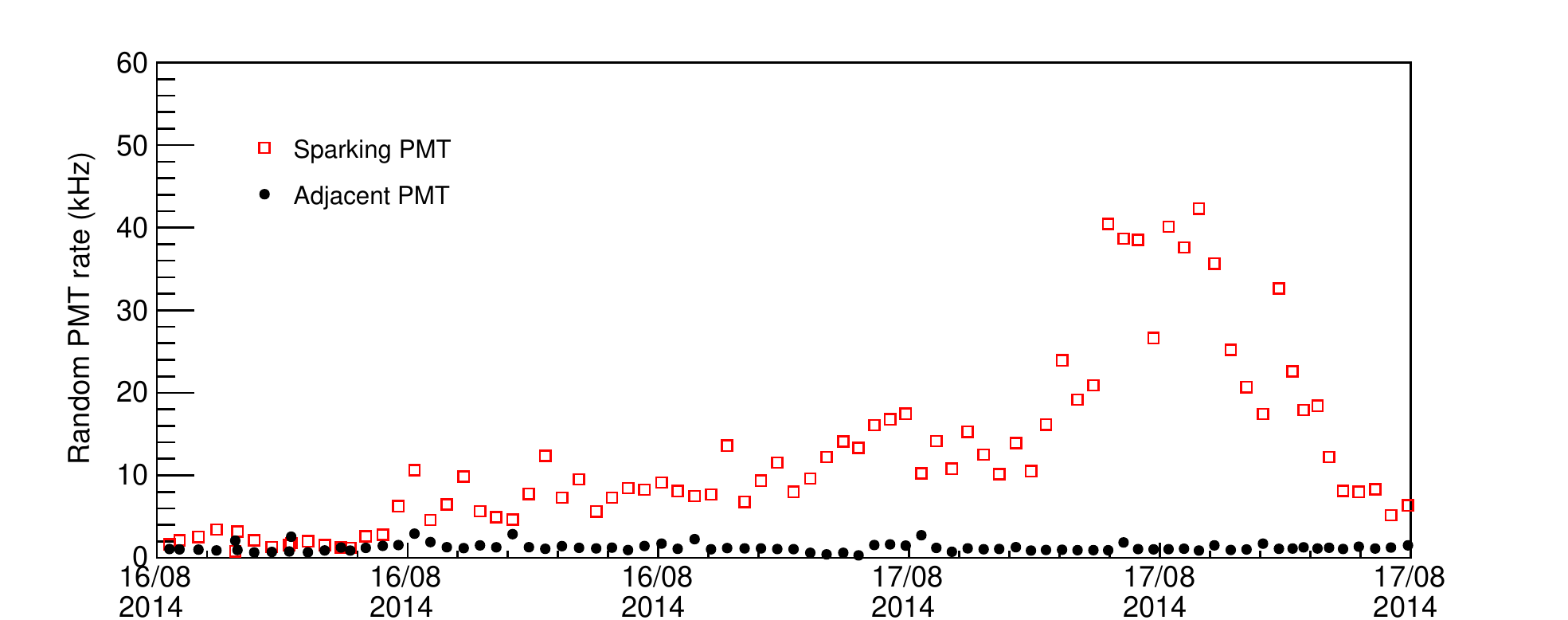}
  \caption{
    Random PMT rate evolution for two neighboring R11410-MOD PMTs; one (red open 
    square) had a sudden increase but 
    no change was observed in the other (black solid circle).
  }
  \label{fig:dark_rate_isolated}
\end{figure}

\begin{figure}[!htbp]
\centering
\subfigure[\it Dark matter data.]
{
\label{fig:dr_rs1_3inch_dm}
  \includegraphics[width=0.8\linewidth]{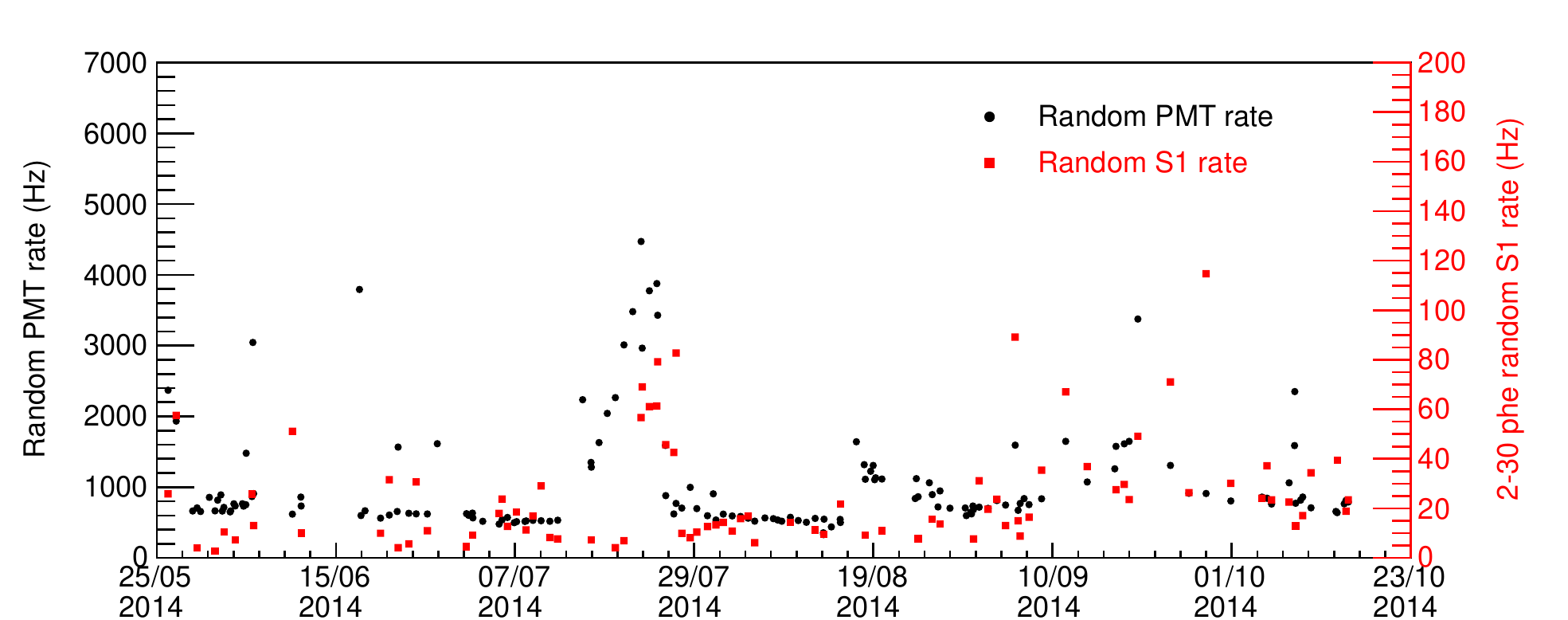}
}
\subfigure[\it $^{60}$Co calibration data.]
{
\label{fig:dr_rs1_3inch_co}
  \includegraphics[width=0.8\linewidth]{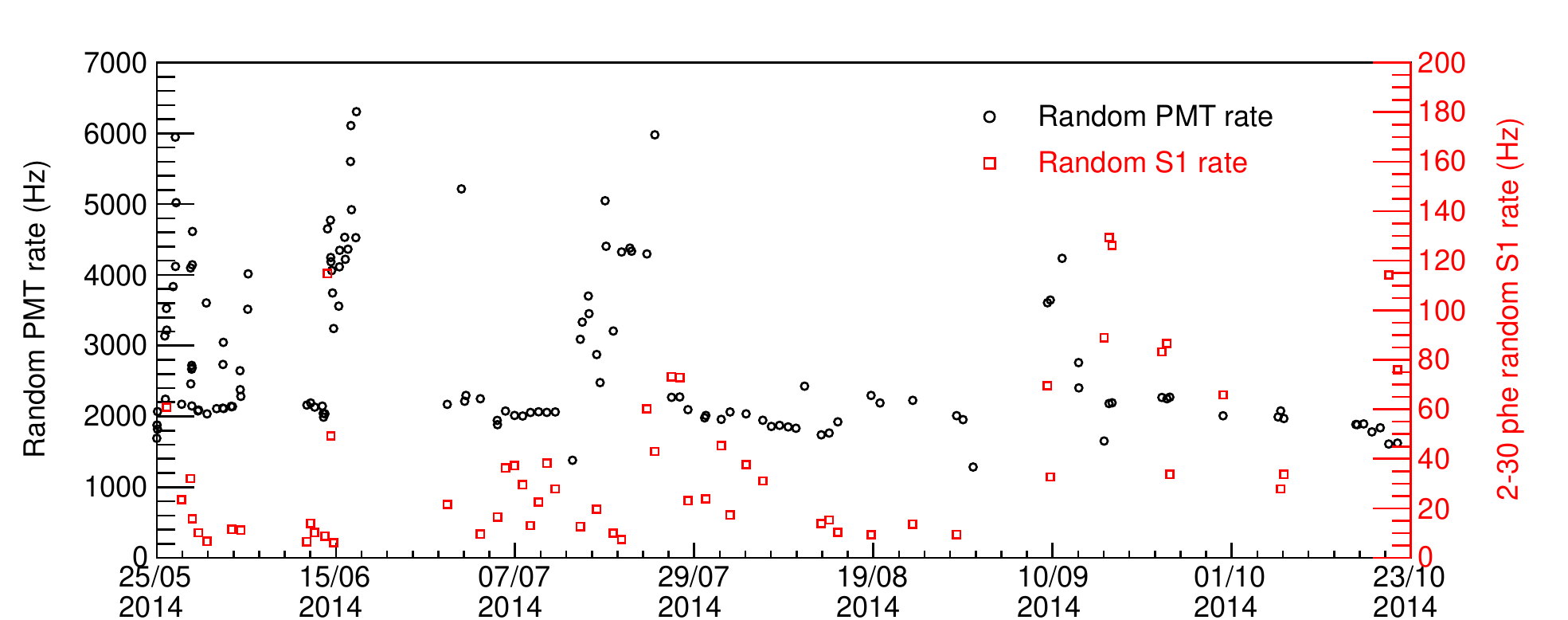}
}
\caption{
    Average random PMT rate and the random S1 rate (indicated by legends) 
    in 2 to 30 PE versus time in dark matter data (a) and $^{60}$Co 
    calibration data (b).
    The vertical scale for the random S1 rates is labeled on the right.
  }
\label{fig:ss_dr_3inch}
\end{figure}

During the operation, the random PMT rates were not stable over time.
Immediately after the xenon fill, the random PMT rate started from hundreds 
of kHz per PMT, but gradually decreased to $\sim$kHz in a few days after the start 
of the liquid purification.
We therefore suspect that the random PMT rate was correlated with the impurity level
since photons from the discharging or S2 electroluminescence could liberate electrons
from impurity atoms, generating additional photons inside the detector. During the stable data
taking period, there were two typical types of rate excursions.
The first type was that the rate increase was observed in an
isolated PMT and no neighboring PMTs were affected, as illustrated 
in Fig.~\ref{fig:dark_rate_isolated}. Such a problem could sometime be cured by 
power-cycling the supply voltage of the affected PMT, indicating that the random PMT rate 
might be due to discharges inside the PMT, similar to that observed in 
Ref.~\cite{paper:Akimov.light-emission}. This was a relatively minor issue as only 
limited number of channels were affected intermittently. 
The second type of excursion affected PMTs more globally.
As shown in Fig.~\ref{fig:ss_dr_3inch}, such excursions were also found to be correlated 
with the increase of the
random S1-like rate, which differs from the random PMT rate by requiring a coincidence of
at least three different PMTs.
The excursion could have multiple origins. 
Spurious small discharges from the TPC electrodes could produce small 
light pulses affecting a cluster or all of the PMTs. 
In addition, if small amount of 
light was produced on a given PMT base, such light could leak into the sensitive 
region producing signals on a number of PMTs. 
Both hypotheses are supported by the fact that some times the instability 
could be cured by power cycling the HV for the electrodes and PMTs.
One also observes in Fig.~\ref{fig:ss_dr_3inch} that 
the average random PMT rate during the $^{60}$Co calibration run is significantly 
higher than that in the dark matter search run, confirming that the 
random PMT rates are not just intrinsic PMT properties but are also sensitive to 
the liquid xenon radiation environment. Given contributions from these 
additional photons generated from the detector, the mild
decrease of random PMT rate at the liquid xenon temperature observed 
in R11410-MOD here is in contrast with the dark rate behaviors reported 
in other bench measurements where more significant decrease
was observed~\cite{paper:Akimov.light-emission, paper:K.Lung, paper:LauraB.}.



\subsection{Afterpulsing}
Electrons emitted from the photocathode ionize the residual gas molecules trapped 
in the dynode structure. Positive ions drift to the PMT cathode under the electric 
field producing additional electrons when impinging onto the photocathode. This produces the so-called 
afterpulses after the primary photoelectron signals. The delayed 
time between the primary and afterpulse is approximately
\begin{equation}
\label{eq:delay_time_func}
\Delta t = \sqrt{\frac{2md}{qV}}\,,
\end{equation}
in which $d$ and $V$ are the distance and voltage between the cathode 
and the first dynode where most of the residual gas molecules are expected,
and $m$ and $q$ are the mass and charge of the ion, respectively.

\begin{figure}[!htbp]
\centering
\subfigure[\it Charge versus delayed time of the afterpulses.]
{
  \label{fig:AP_q_vs_t}
  \includegraphics[width=.45\linewidth]{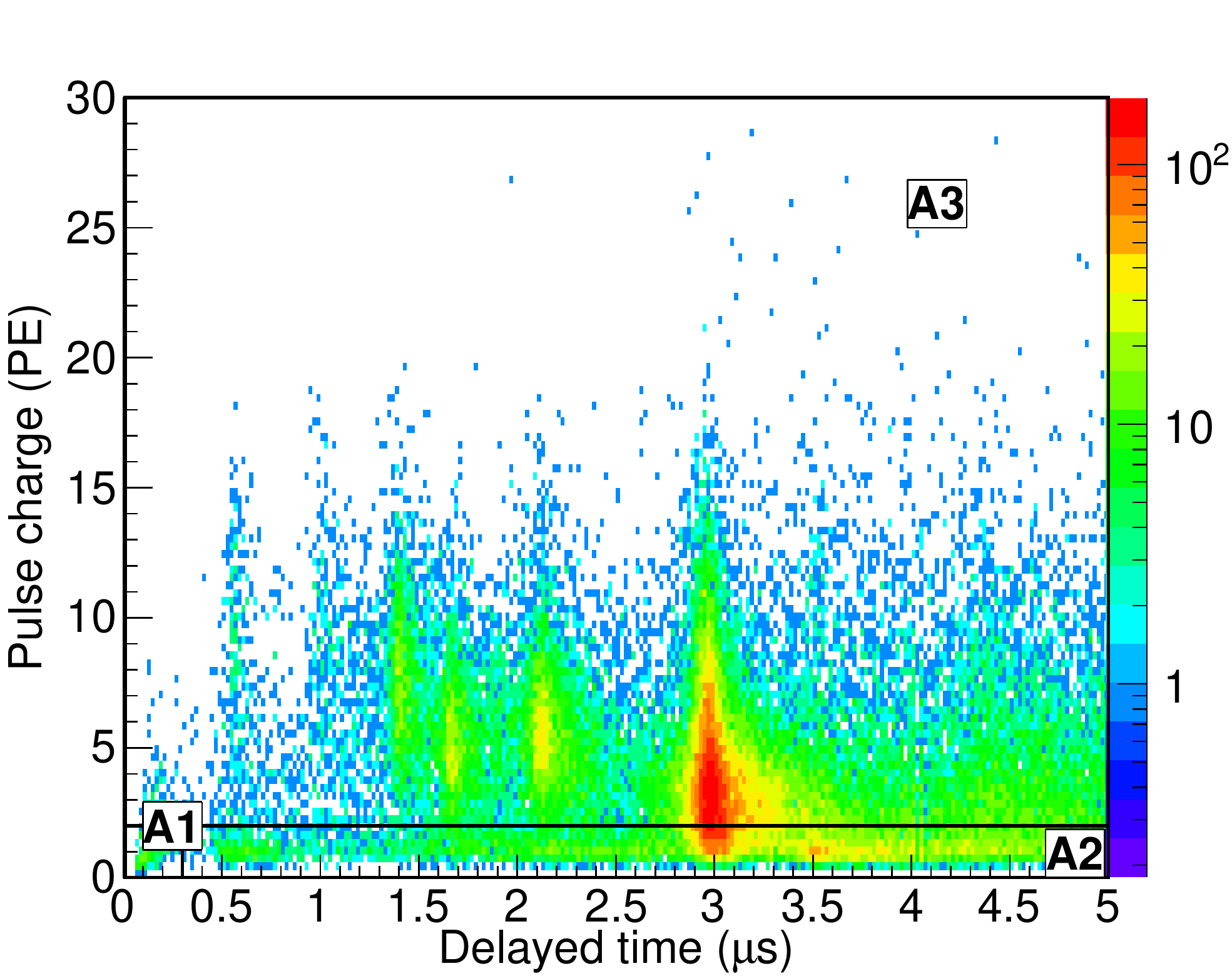}
}
\subfigure[\it Delayed time distribution.]
{
  \label{fig:AP_t}
  \includegraphics[width=.45\linewidth]{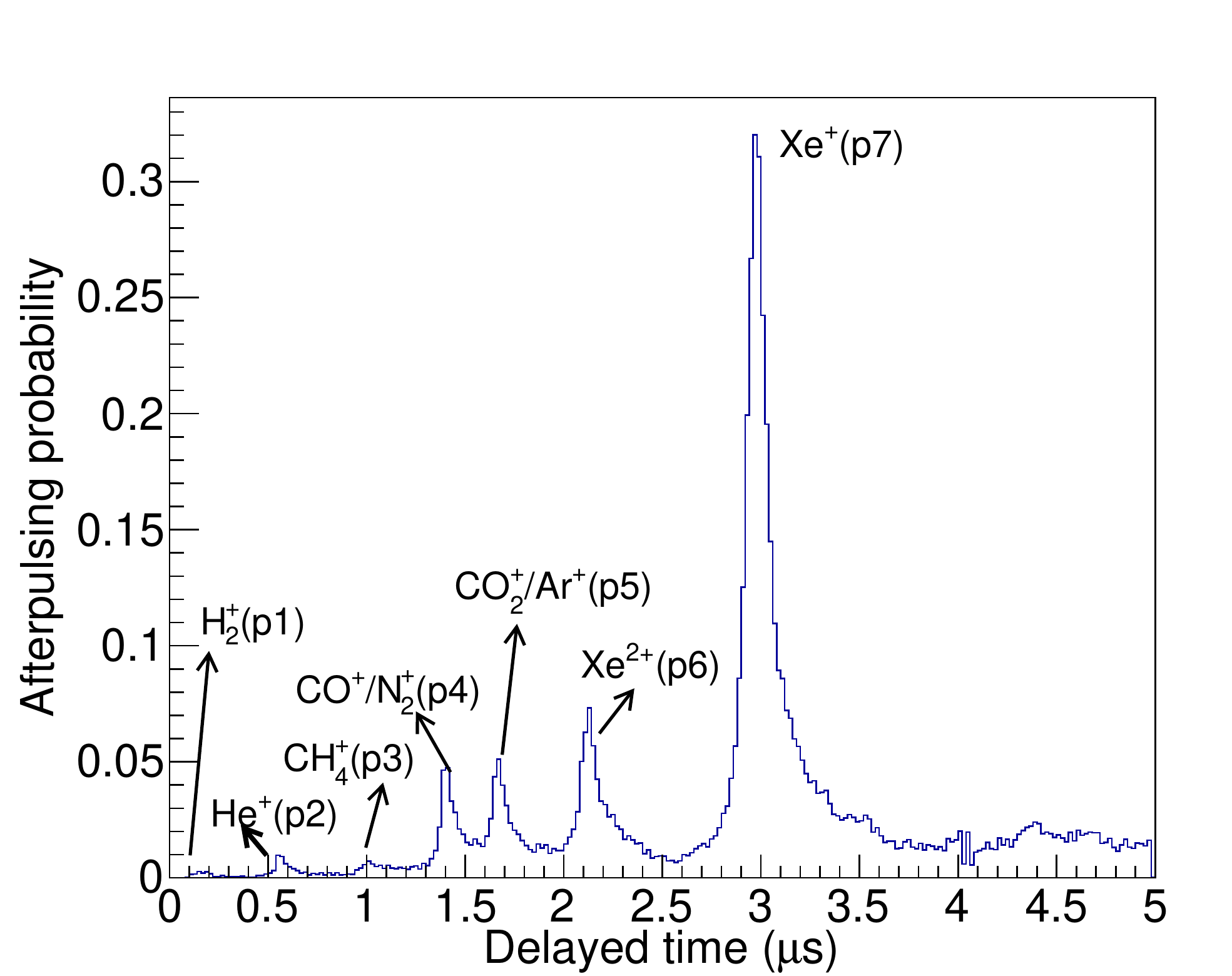}
}
\subfigure[\it Charge distribution.]
{
\label{fig:AP_q}
  \includegraphics[width=.45\linewidth]{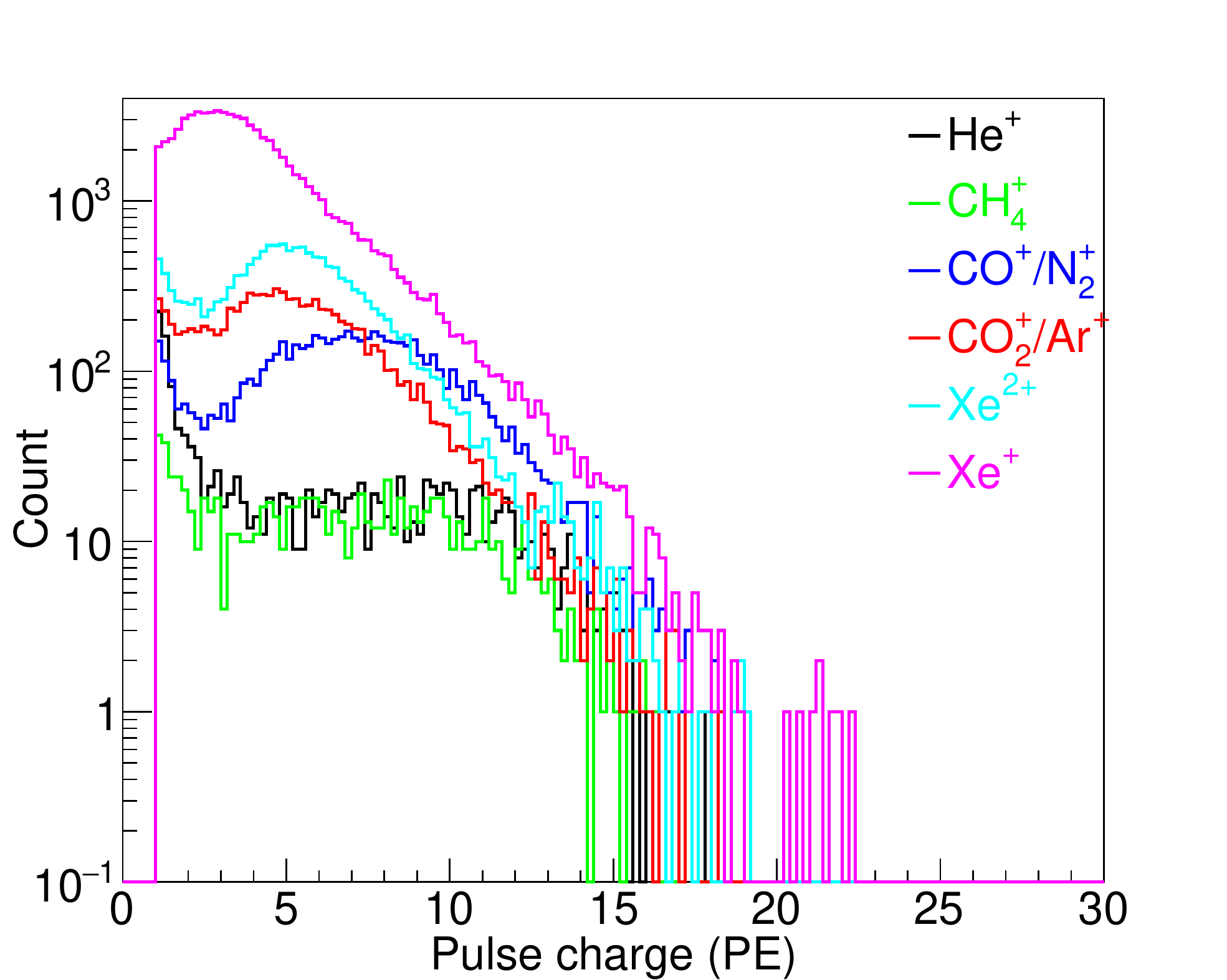}
}
\caption{Distributions of the afterpulses for a R11410-MOD PMT with high afterpulsing
rate: (a) charge vs. delayed time, each point representing an afterpulse
identified after the primary pulses $<20$PE; (b) the delayed time distribution 
of the afterpulses in the A3 region with individual ionic components indicated; and (c) 
charge distribution of the afterpulses associated with different ionic components (see 
legend). 
}
\label{fig:AP}
\end{figure}

Previous bench measurements of the afterpulsing for R11410 PMTs have been 
performed by number of 
groups~\cite{paper:K.Lung, paper:LauraB., paper:Akimov.noise-characteristics}.
In PandaX-I data, afterpulses were identified {\it in situ} during 
data taking by selecting pulses within 5~$\mu$s after a 
small primary pulse with charge $<$20 PE. For each afterpulses, 
the delayed time and the charge of the afterpulse was recorded.
For illustration, the two-dimensional distributions of delayed time vs. charge 
as well as the respective one-dimensional projections are 
shown in Fig.~\ref{fig:AP} for a R11410-MOD PMT with 
relatively high afterpulsing rate.  
Using the same classification as in Ref.~\cite{paper:LauraB.}, 
Fig.~\ref{fig:AP_q_vs_t} was separated into three
regions as indicated. A1 represents small pulses ($\leq$2 PE) happened within 
0.3~$\mu$s after the primary hit. 
The delay is mostly caused by elastic scattering of secondary 
electrons on the first dynode~\cite{Hamamatsu}.  
A2 represents random hits ($\leq$2 PE) 
between 0.3 to 5~$\mu$s after the primary hit. A3 contains the 
afterpulses ($>$2 PE) caused by the residual gas ions.
Peaks due to different ions can be clearly identified from the timing distribution in 
Fig.~\ref{fig:AP_t}, similar to those in 
Refs.~\cite{paper:K.Lung, paper:LauraB., paper:Akimov.noise-characteristics}.
Peak located at around 3 $\mu$s is identified as the Xe$^{+}$ peak, and similarly all other 
peaks are identified using Eqn.~\ref{eq:delay_time_func}. 
The charge distributions of the afterpulses for
individual ionic peaks in Fig.~\ref{fig:AP_t} are shown in Fig.~\ref{fig:AP_q}. 
For all $+1$ charged ions, 
one finds qualitatively
that the average charge of the afterpulses grows with the decrease of the 
mass of the ion, as expected from their increasing ionization capability.

\begin{figure}[!htbp]
\centering
  \includegraphics[width=.7\linewidth]{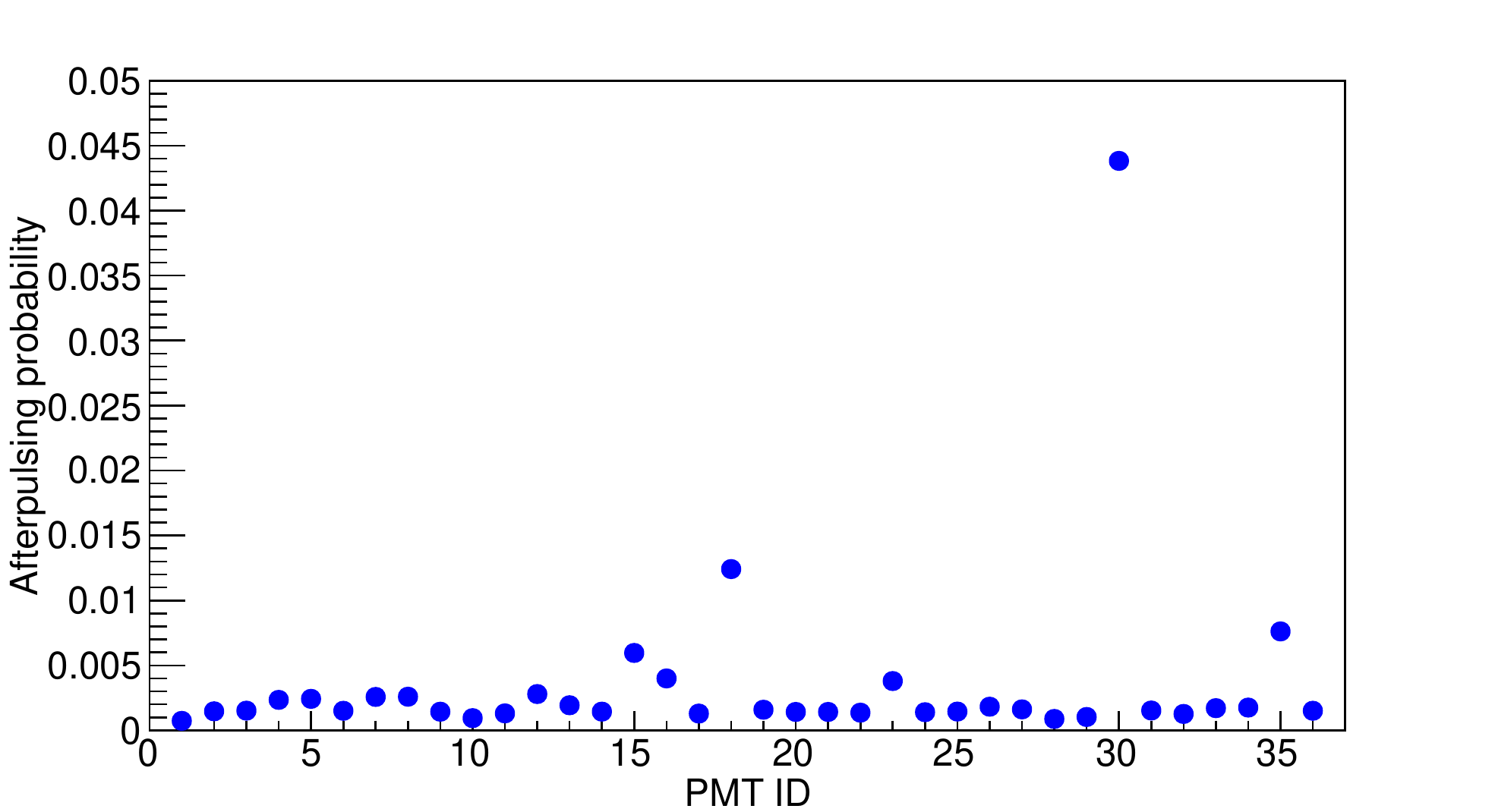}
\caption{Measured APPs for individual active R11410-MOD PMTs.}
\label{fig:APR_area3}
\end{figure}

The probability of occurrence of afterpulses per single PE (APP) in a given data run is
\begin{equation}
\label{eq:apr_func}
APP = \frac{\sum N_{afterpulse}}{\sum Q_{primary}}\,,
\end{equation}
where $\sum N_{afterpulse}$ is the total number of identified afterpulses and 
$\sum Q_{primary}$ 
is the summation of the charge of the primary in the unit of PE.
Since the primaries were selected to be $<20$PE, pile-up of afterpulsing 
per primary is negligible, and the APP in Eqn.~\ref{eq:apr_func} 
was verified to be insensitive to the cut 
to the primary charge.
In Fig.~\ref{fig:APR_area3}, the APP in A3 (Fig.~\ref{fig:AP_q_vs_t})
for all active R11410-MOD PMTs are summarized. 
One sees that the APPs for most of the 
PMTs are less than 1\%, except a few which are at a few percent level. 
The average APP for the R11410-MOD PMTs is 1.7\%,
with details listed in Tabel~\ref{tab:R11410apr}.
We observe no correlation between the APP and the measured random PMT rate. 
The average APP for the R8520-406 tubes is less than 0.3\%. 

\begin{table}[!htbp]
\centering
\begin{tabular}[c]{l|c|ccc|ccccccc}
\hline\hline
   PMT & Total (\%) & A1 & A2 & A3 & p1 & p2 & p3 & p4 & p5& p6 & p7 \\
\hline
   AVE & 1.70 & 0.38 & 1.13 & 0.19 & 0.05 & 0.03 & 0.01 & 0.02 & 0.03 & 0.02 & 0.02 \\
\hline
ZK6307 & 4.34 & 0.67 & 2.43 & 1.24 & 0.11 & 0.06 & 0.01 & 0.04 & 0.24 & 0.14 & 0.65 \\
ZK6313 & 8.97 & 0.03 & 4.56 & 4.38 & 0.01 & 0.05 & 0.04 & 0.28 & 0.37 & 0.64 & 2.99 \\
KA0019 & 2.49 & 0.44 & 1.30 & 0.76 & 0.06 & 0.03 & 0.01 & 0.01 & 0.11 & 0.11 & 0.44 \\
\hline\hline
\end{tabular}
\caption{Summary of average R11410-MOD PMT APPs 
as well as those for the abnormal ones. The entries p1 to p7 in the table refer to the 
APPs for H$_2^+$, He$^{+}$, 
CH$_4^+$, CO$^+$/N$_2^+$, CO$_2^+$/Ar$^+$, Xe$^{2+}$,
and Xe$^{+}$ peaks, respectively.}
\label{tab:R11410apr}
\end{table}

\begin{figure}[!htbp]
\centering
\subfigure[\it A typical PMT.]
{
  \includegraphics[width=.8\linewidth]{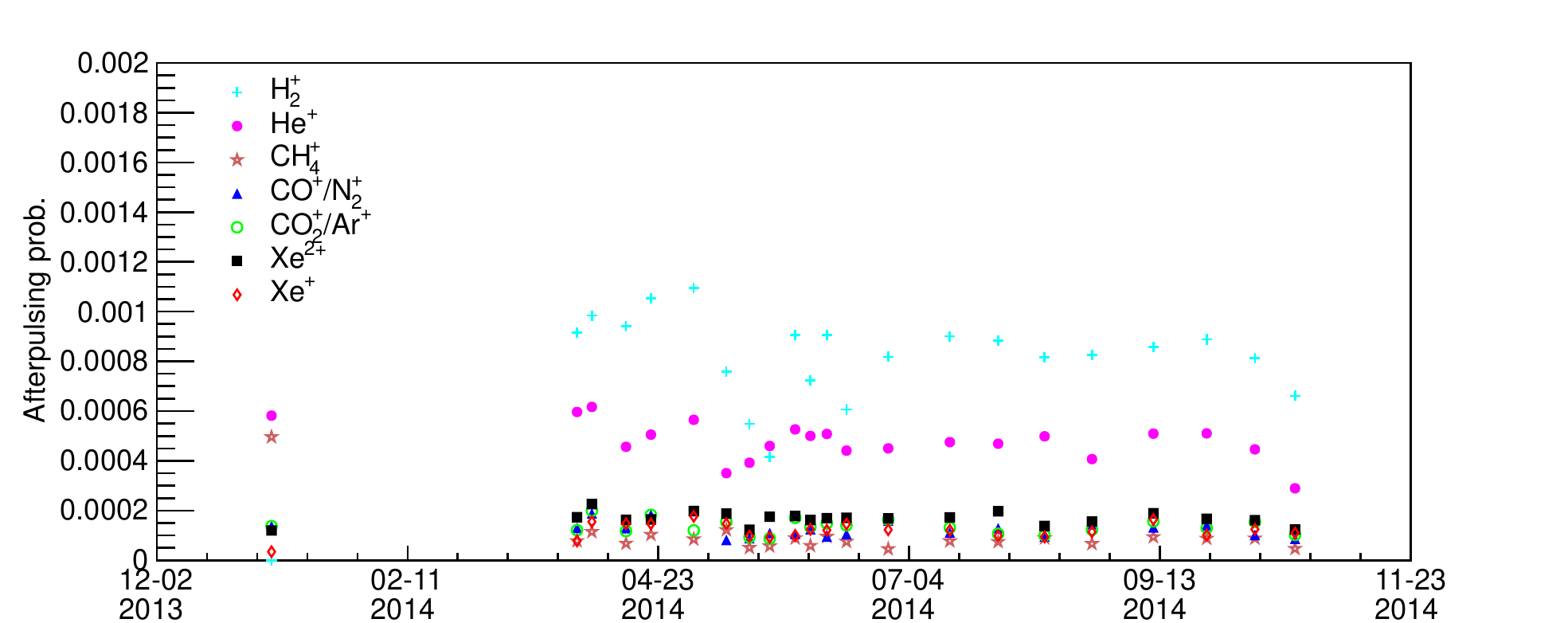}
}
\subfigure[\it PMT with abnormal APP.]
{
  \includegraphics[width=.8\linewidth]{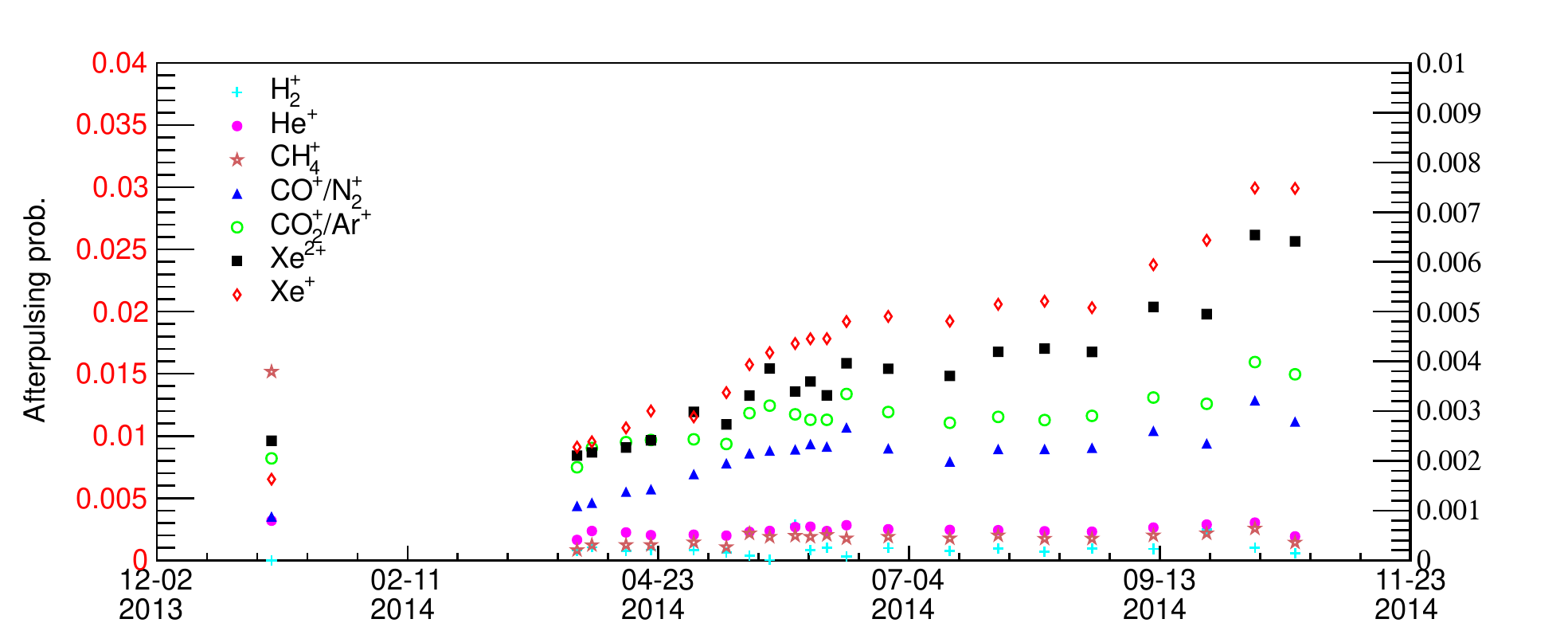}
}
\caption{
APP vs. time for two R11410-MOD PMTs, one typical (a) and one abnormal (b). 
The very first data point in each figure is obtained in the gas xenon run under 
room temperature. Note that in (b) the scale for Xe$^{+}$ is indicated on the left side and
the others on the right. 
}
\label{fig:APR_history}
\end{figure}

The evolution of the APPs from two PMTs, one typical, the other one abnormal, 
is shown in Fig.~\ref{fig:APR_history}. 
The typical one is rather stable over time. 
For the abnormal PMT, the APP of all ionic components 
started to increase from the beginning of the liquid xenon run, 
strongly indicating a leak related effect. 

\subsection{Saturation for large pulses}
\begin{figure}[!htbp]
\centering
  \includegraphics[width=.5\linewidth]{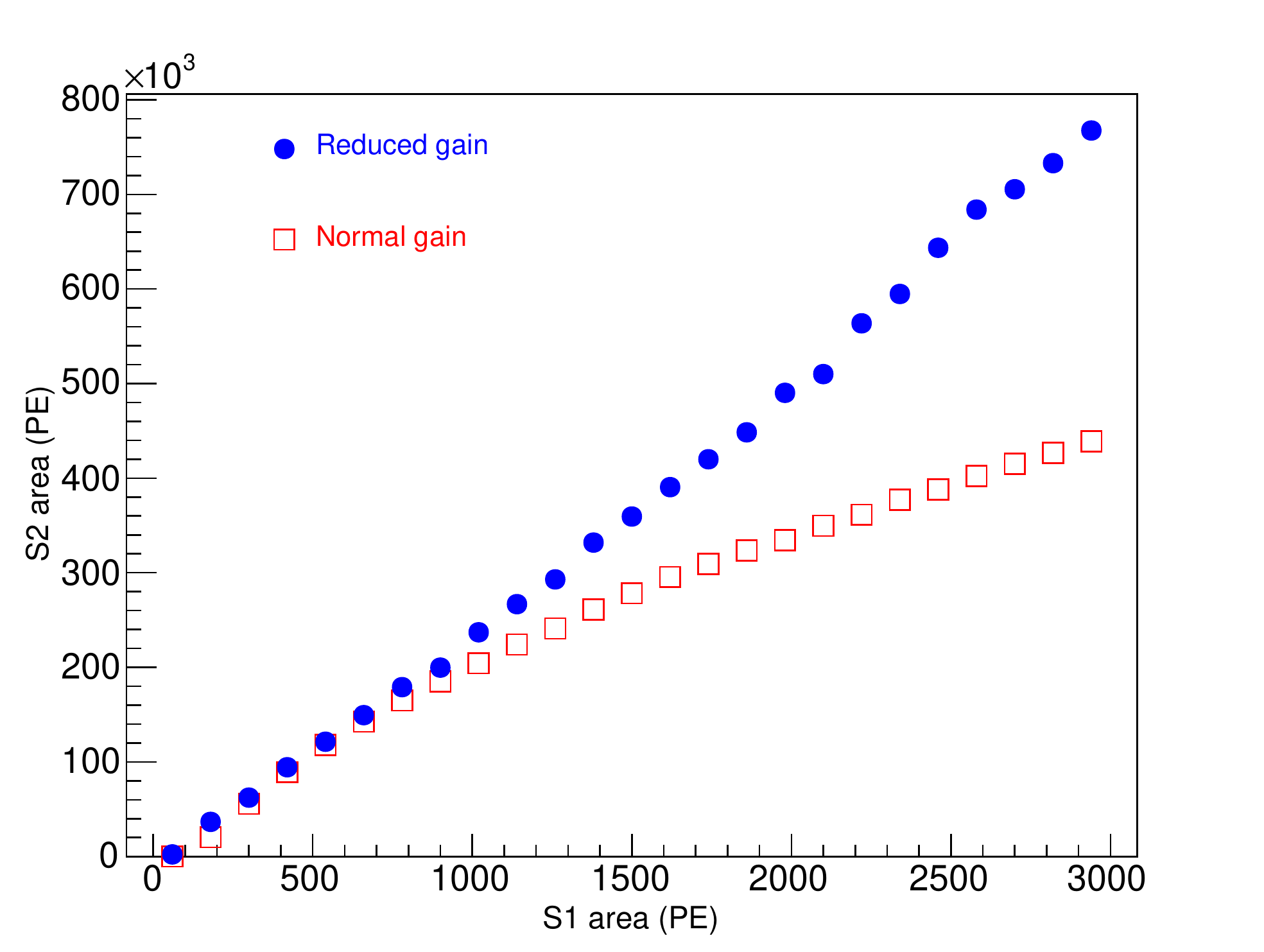}
\caption{The mean S2 vs. S1 in PE when all PMTs were operated at the normal gain 
of 2$\times 10^{6}$
(red) and reduced gain of  5$\times 10^{5}$ (blue). Saturation of S2 signals was observed
in the normal gain data for high energy events. 
The reduced gain data showed improved linearity for larger S2, but 
had a slight ``overshoot'' for S1 around 200 PE 
due to a higher trigger threshold for S2, as well as a lower signal finding efficiency for S1.}
\label{fig:normal_reduced_area_profileX}
\end{figure}

%
As mentioned in Sec.~\ref{sec:base}, to reduce the radioactivity level from the 
ceramic capacitors, we only kept one capacitor between the last dynode and the anode.
This was not a problem for the low energy signals in the dark matter search 
region, but the linearity of the PMT for large pulses, S2 pulses in particular, 
was affected. On the other hand, 
if the supply voltage was reduced, the linearity could be recovered to a certain extent.  
To demonstrate this, in Fig.~\ref{fig:normal_reduced_area_profileX} the measured 
S2 charge vs. S1 charge are shown for two different PMT high voltage settings. 
One sees that
under the normal HV setting, the measured S2 starts to exhibit saturation for S1 above 
$\sim$600 PE which is approximately 194 keV$_{ee}$ electron-equivalent energy. 

\subsection{Other performance issues}
During the PandaX-I operation, 
one bottom PMT was fully disabled due to unstable random PMT rate.
We suspect that intermittent discharging 
happened on the PMT base assembly (Sec.~\ref{sec:dark_rate}). In addition,
four bottom and eight top PMT channels gradually developed
connection problems, resulting in improper base resistance, capacitance, or 
frequent HV trips, and were disconnected. 
To avoid complication due to a time-dependent light yield, in the full exposure analysis 
we inhibited all these 13 PMTs.
This, and the updated baseline calculation algorithm~\cite{pandaxI2nd}, are the two
major factors contributing to the decrease of average light yield
for S1 from 7.3 PE/keV$_{ee}$ (standard zero electric field value at
122 keV$_{ee}$) in the first 17.4-day data period~\cite{pandaxI1st} to 
6.0 PE/keV$_{ee}$ in the full exposure data~\cite{pandaxI2nd}.   
Since the bottom PMTs dominated the light collection for S1, and the photon 
distribution was measured to be rather uniform independent 
of the interaction vertex, this represents a constant downward correction to the photon 
detection efficiency. Using a dedicated Monte Carlo including the photon propagation 
in the TPC, the effects due to disabled top PMTs to the position reconstruction were studied 
and verified to be minor. 
To set the scale, at the 300 PE S2 event selection threshold~\cite{pandaxI2nd},
for events located at the center of a disabled PMT, reconstructed position resolution
changed from about 5 mm to less than 7 mm, leading to a negligible influence to the
fiducial volume selection.



\section{Summary}
We report the long term performance of 180 PMTs in the
PandaX-I experiment operated for nine months in a liquid xenon environment.
Details of key PMT parameters such as the gains, SPE resolutions, 
random PMT rates, and afterpulsing probabilities are presented. 
The unstable random PMT rate was the most important issue identified during the 
run. The problem appeared to be related to the discharges either inside the PMT, 
causing excessive random PMT rates in isolated PMTs, 
or from the base assemblies and/or the TPC electrodes, affecting a cluster or all PMTs. 
This study provides an important guidance to the design of the PandaX-II experiment, 
as well as future xenon-based 
dark matter experiments using R11410 series as the primary photon sensors.

\section{Acknowledgement}
The PandaX project has been supported by a 985-III grant from Shanghai Jiao Tong
University, a
973 grant from Ministry of Science and Technology of China (No. 2010CB833005),
and grants from National Science Foundation of China (Nos. 11055003, 11435008, 
11455001, and 11525522).
This work is supported in part by the 
Shanghai Key Laboratory for Particle Physics and Cosmology, 
Grant No. 15DZ2272100, and the CAS Center for Excellence in Particle Physics.
The work has also been sponsored by Shandong University, Peking University,
and the University of Maryland.







\end{document}